\documentclass[letterpaper,twocolumn,10pt]{article}
\usepackage{usenix2023_SOUPS}

\usepackage{longtable}
\usepackage{fontawesome5}
\usepackage{booktabs}
\usepackage{multirow}
\usepackage{array}
\usepackage{xcolor}
\usepackage{colortbl}
\usepackage{xspace}
\usepackage{mathtools}
\usepackage{xurl}
\usepackage{hyperref}
\usepackage{balance}

\begin{document}

\date{}

\title{\Large \bf SoK: Analysis of User-Centered Studies Focusing on Healthcare Privacy \& Security}

\def\plainauthor{Author name(s) for PDF metadata. Don't forget to anonymize for submission!}

\author{ 
{\rm Faiza Tazi$^1$, Archana Nandakumar$^2$, Josiah Dykstra$^3$, Prashanth Rajivan$^2$, Sanchari Das$^1$}\\
University of Denver$^1$, University of Washington$^2$, Designer Security$^3$
} 

\maketitle
\thecopyright

\begin{abstract}
Sensitive information is intrinsically tied to interactions in healthcare, and its protection is of paramount importance for achieving high-quality patient outcomes. Research in healthcare privacy and security is predominantly focused on understanding the factors that increase the susceptibility of users to privacy and security breaches. To understand further, we systematically review $26$ research papers in this domain to explore the existing user studies in healthcare privacy and security. Following the review, we conducted a card-sorting exercise, allowing us to identify $12$ themes integral to this subject such as ``Data Sharing,'' ``Risk Awareness,'' and ``Privacy.''  Further to the identification of these themes, we performed an in-depth analysis of the $26$ research papers report on the insights into the discourse within the research community about healthcare privacy and security, particularly from the user perspective.
\end{abstract}

\section{Motivation}

Security and privacy integration in the healthcare domain is essential to protect patients' data~\cite{dwivedi2019decentralized}, considering medical records include sensitive health and personal information. The healthcare industry is often a prime target for cybercriminals considering that these data sets could contain a plethora of sensitive information such as social security numbers, birth dates, employment information, emergency contacts, and insurance and billing data; these data are also notoriously difficult to monitor or safeguard after a breach~\cite{kruse2017cybersecurity}. Furthermore, healthcare data are lucrative on the black market. Sahi et al. noted that sensitive medical data are sold for an average of \$40-50 per record~\cite{sahi2017privacy}. In light of this and to understand what is studied on the healthcare data privacy and security from the user side in research literature, we conducted a systematic literature review. 
\section{Method}
\label{sec:method}
We conducted a systematic literature review including a corpus of $129$ papers published up to December 10, 2021 of user studies with a focus on privacy and security of healthcare patients' data. Papers were excluded if they were presented as a work-in-progress (posters, extended abstracts, less than $4$ pages long, etc.). We collected papers from seven digital databases: ACM Digital Library (DL), Google Scholar, SSRN, ScienceDirect, IEEE Xplore, PubMed, and MEDLINE. After the initial search to obtain the keywords, we collected the papers using keywords like ~\textit{Healthcare Data Security, Healthcare Data Breach, Healthcare Data Theft, Medical Data Theft, Medical Data Security, Medical Data Breach, Patient Data Security, Patient Data Theft, and Patient Data Breach} through the Publish or Perish~\footnote{https://harzing.com/resources/publish-or-perish} software for retrieving articles from Google Scholar. After removing any duplicate articles we were left with $129$ papers. We adapted the study design from prior systematic reviews~\cite{tazi2022sok,majumdar2021sok,das2022sok,shrestha2022sok,stowell2018designing,duzgun2022sok,noah2021exploring,das2019all,das2019evaluating}.

After analyzing the full text of the $129$ papers, we excluded $49$ papers from the set because the works though mentioned healthcare and the concerns of the data from the privacy and security lenses as a motivational factor were not directly focused on privacy and security of healthcare data. From the remaining $n=80$ papers, we consolidated the papers which consistently addressed healthcare data privacy and security throughout various stakeholders' perspective. We were left with $26$ papers on which we conducted a card-sorting exercise involving all authors.

\section{Results}
\textit{Risk perception:} It is challenging to circumscribe the perception of risk as risks do not have the same meaning for everyone. That is why user studies focusing on risk perception are critical, especially for the subject. Papers were categorized in the risk perception label when part of the study or its entirety explored participants' attitudes, and opinions on risks related to healthcare data. Risk perception was the most frequent label in our corpus where $61.54\%$ of the papers were within this category.
\\
\textit{Data sharing:} $14$ papers aimed to understand the perspective of participants on data sharing practices that would be acceptable to patients and beneficial to research communities. Results from these papers indicating that patients support data sharing if it benefits the public, or if the data is shared for personal health purposes. Nonetheless, people still have reservations about the privacy of sensitive data, data breaches, and medical bias.
\\
\textit{Electronic Health Records (EHR):} We found eight papers in our corpus pertaining to user interactions with EHR. These papers confirm through their results that participants have concerns over privacy and security, and are prudent about using EHR technologies. It was also determined that providers' reassurance positively impact patients' continuous and systematic usage of patient portal software in general and lowers their security concerns.
\\
\textit{Risk Awareness:} Despite the abundant potentialities for cyber risk in the healthcare sector~\cite{ahmed2017false}, there is a startling level of naiveté among some healthcare providers. The results from the $8$ papers relevant to risk awareness, show that the knowledge levels of  providers regarding patient privacy, confidentiality and data sharing practices is average or lower. 
\\
\textit{Technology Adoption:} 
Technology adoption in the healthcare domain is crucial to its development. To this regard, eight papers in our corpus examined factors and inspected participants' requirements to improve user acceptance and adoption of some healthcare technologies. The results reported by these papers reveal that the security and privacy aspects bolster the acceptance and adoption of healthcare technologies. 
\\
\textit{Regulatory Compliance:} Seven papers studied the ethical and legal aspects of healthcare data management. These papers mainly assess the HIPAA compliance of participants, as well as the cybersecurity conditions and behavior of healthcare practitioners and organizations. According to the CDC, ``The Health Insurance Portability and Accountability Act of 1996 (HIPAA) is a federal law that required the creation of national standards to protect sensitive patient health information from being disclosed without the patient’s consent or knowledge''~\cite{CDCHIPAA}. Notably all the studies here determined that there needs to be more policies and reinforcement of behavior which can impede security.
\\
\textit{Individual Differences:} Comparisons can be based on experience level, hospital size, marriage status, country of origin, health status, or gender. We found seven  papers from our corpus who did this type of analysis. In particular, Wilkowska and Ziefle show that females and healthy adults demand the highest security and privacy standards compared to males and the ailing elderly~\cite{wilkowska2012privacy}. A different study, 
investigated the extent to which security policies impact health information interoperability at different levels within the same hospitals~\cite{shrivastava2021data}.
\\
\textit{Secure Communications:} In the case of healthcare, secure communications is not just a matter of security and privacy, but it can also be a medical concern. 
We categorized five papers within this label. Most of these papers have results that show that patients still do not fully trust the existing communications technologies, except for Elger's study~\cite{elger2009violations} where $85\%$ of the participants had no privacy concerns regarding using a secure SMS system for private medical communications.
\\
\textit{Mobile Applications:} Three papers were related to mobile applications. These papers evaluate users' perceptions of mobile health apps regarding privacy, security, and quality of care. The results of these papers were somewhat different, where Schnall et al.~\cite{schnall2015trust} found that the majority of their participants were concerned over privacy of their sensitive healthcare data and people having access to their healthcare data. On the other hand, both Giguere et al.~\cite{giguere2018participants} and Richardson and Ancker's~\cite{richardson2015public} studies found that the majority of participants are unconcerned about privacy when using such apps.
\\
\textit{Social Influence:} 
Three papers in our corpus were categorized as social influence. These papers proved that participants were influenceable. Namely, Moqbel et al.~\cite{moqbel2021sustaining} demonstrated that health professionals' reassurance and encouragement positively impact patients' continuous and systematic usage of patient portal software; not only that but participants were also influenced to lower their security concerns through the same encouragement. 
\\
\textit{Privacy:} 
Most of the papers in our corpus touch upon privacy, but three of these papers were directed exclusively towards the privacy of healthcare data. Accordingly, in their study Elger~\cite{elger2009violations} assesses the knowledge and perceptions of physicians on healthcare data violations of privacy; results show that 11\% of the participants recognized all the confidentiality violations in the test cases they were presented with.
\\
\textit{Contact Tracing:} Only two papers were categorized as contact tracing. 
With the emergence of digital contact tracing applications, users have expressed privacy and security concerns~\cite{chan2021privacy}. These concerns stem from apprehension of data breaches or having their data collected by government entities~\cite{hassandoust2021individuals}. However, this did not deter participants from approving COVID-19 contact tracing apps. 
\\
We have provided the details of the papers and the themes including the snapshot in the card sorting exercise in the Appendix~\ref{appendixA}.

\bibliographystyle{plain}
\bibliography{usenix2023_SOUPS}
\balance
\newpage
\onecolumn
\appendix
\begin{figure}
    \centering
    \includegraphics[width=0.9\textwidth]{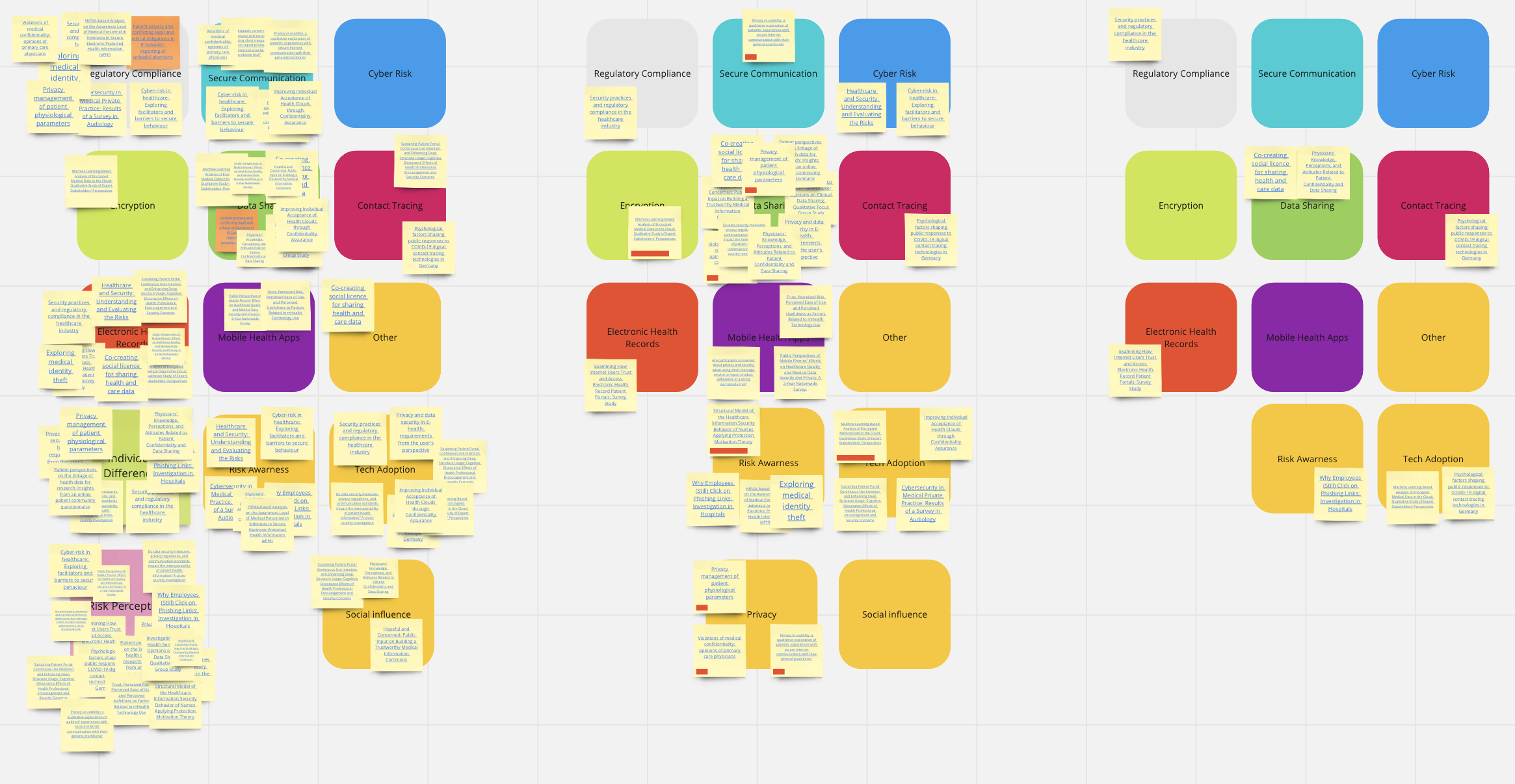}
    \caption{A snapshot of the card-sorting exercise by the researchers of the paper used to analyze the paper repository.}
    \label{fig:cardsort}
\end{figure}

\section{Overview of Security-Focused User Studies}
\label{appendixA}
\begin{center}
\begin{small}
\begin{longtable}{|p{0.7cm}|p{3.6cm}|p{3.4cm}|p{5.5cm}|p{0.9cm}|}

\hline \multicolumn{1}{|c|}{\textbf{Study}} & \multicolumn{1}{c|}{\textbf{Goal}} & \multicolumn{1}{c|}{\textbf{Methods}}& \multicolumn{1}{c|}{\textbf{Principal Findings}} & \multicolumn{1}{c|}{\textbf{Labels}} \\ \hline 
\endfirsthead

\multicolumn{3}{c}%
{{\bfseries \tablename\ \thetable{} -- continued from previous page}} \\
\hline \multicolumn{1}{|c|}{\textbf{Study}} & \multicolumn{1}{c|}{\textbf{Goal}} & \multicolumn{1}{c|}{\textbf{Methods}} & \multicolumn{1}{c|}{\textbf{Principal Findings}}& \multicolumn{1}{c|}{\textbf{Labels}}\\ \hline 
\endhead

\hline \multicolumn{3}{|r|}{{Continued on next page}} \\ \hline
\endfoot

\hline \hline
\endlastfoot

    \hline
    \cite{jalali2020employees} & Understand reasons why hospital employees click on phishing emails & Quantitative: partial least squared structural equation modeling & 
        Workload has a significant negative effect on secure behavior & \faUserFriends  \newline \faExclamationCircle \newline \faExclamationTriangle\\
    \hline
    \cite{giguere2018participants} & Assess participants' attitudes towards privacy and security while using system developed for a medical study & Mixed Methods: descriptive statistics + analysis of variance for quantitative data, thematic analysis for qualitative data  & The majority of participants are unconcerned about privacy and confidentiality when using SMS despite the fact that some participants expressed their concern about possible data leaks & \faExpeditedssl \newline \faShareSquare
    \newline \faExclamationTriangle \newline \faMobile\\
    \hline
    \cite{dykstra2020cybersecurity} & Assess HIPAA compliance, cybersecurity conditions and behavior of healthcare practitioners in private practices  & Quantitative: descriptive statistics & $9.9\%$ of the participants confirm they experienced at least one data breach in 2019 \newline $24.4\%$ participants claim they have cyber insurance   & \faGavel \newline \faExclamationCircle \newline \faUsb\\
    \hline
    \cite{kwon2013security} & Assess security practices of healthcare organizations & Quantitative:  Ward's cluster analysis using minimum variance & 
    Participating hospitals were clustered into three clusters:leaders, followers, and laggers \newline
    Hospitals prioritize technical security solutions and data privacy over security management processes and performing regular audits & \faGavel \newline \faMedrt \newline \faUserFriends \newline \faUsb \newline \faExclamationTriangle\\
    \hline
    \cite{lee2021structural} & Assess nurses' health information security (HIS) practices & Quantitative: exploratory + confirmatory factor analysis & The participant nurses' HIS intentions are affected by the amount of HIS losses they are able to handle``coping appraisal'' ($estimate=-1.477$, $p<0.01$) \newline HIS intentions have a considerable impact on coping appraisal ($estimate=0.515$, $p<0.001$) & \faExclamationCircle \newline \faExclamationTriangle  \\
    \hline
    \cite{masood2018privacy} & Evaluate the extent to which access to patients' physiological parameters (PPP) in hospitals can infringe on the patients' privacy  & Quantitative: bivariate analysis & Patients need to have control over their own PPPs\newline Specialists are the more trusted than family doctors, nurses, and medical assistants & \faGavel \newline \faShareSquare \newline \faUserFriends \newline \faEyeSlash \\
    \hline
    \cite{karasneh2021physicians} & Evaluate physicians' perceptions and understanding of confidentiality and medical data sharing & Quantitative: Pearson's correlation + Multiple regression & Physicians' mean score for knowledge regarding patient confidentiality and data sharing is $7.34$ out of $14$ and is positively correlated with their attitudes towards the subject matter which leads to privacy breaches   &\faShareSquare \newline \faUserFriends \newline \faExclamationCircle \newline\faUsers\\
    \hline
    \cite{wilkowska2012privacy} & Evaluate users' attitudes towards privacy and security of medical technology& Mixed Methods: One-way ANOVA + F-Tests + Spearman's rank correlations for quantitative data and thematic analysis  for qualitative data & Participants with better health value privacy and security  of medical technologies and control over data access more than participant with poor health & \faShareSquare \newline \faUserFriends \newline \faUsb \newline\faExclamationTriangle\\ 
    \hline
    \cite{richardson2015public} & Evaluate the perceptions of users of mobile health applications regarding privacy, security and quality of care & Quantitative: multivariable logistic models + bivariate analysis & In 2014 participants were more likely to think that mhealth improves the quality of healthcare, however they were just as concerned about privacy in 2013 (74\%) as in 2014 (75\%) & \faShareSquare \newline \faMedrt \newline\faExclamationTriangle	 \newline \faMobile\\
    \hline
    \cite{alaqra2021machine} & Evaluate the perceptions of experts on using ML based privacy enhancing technologies (PETs) that enable automated analysis of encrypted healthcare data stored in the cloud  & Qualitative: thematic analysis & Technical experts admonish prudence in trusting ML based PETs \newline Medical experts call for patient safety assurances regarding these tools & \faShareSquare	\newline \faMedrt \newline \faUsb \\
    \hline
    \cite{shrivastava2021data} & Investigate the extent at which security policies impact health information interoperability at different levels within the same hospitals & Quantitative: logistic models & Hospitals with access control implemented in workstations are 44\% less likely to encounter technical interoperability (TI) issues. \newline Hospitals using one EMR are 53\% less like to encounter TI issues compared with hospitals using numerous EMR systems & \faShareSquare \newline \faUserFriends \newline \faUsb \newline \faExclamationTriangle \\
    \hline
    \cite{moqbel2021sustaining} & Assess the influence of healthcare providers' encouragement and patient security concerns in patient portal software continued usage & Quantitative: partial least squares structural equation modeling & Providers' reassurance and encouragement has a positive impact on patients' continuous use and systematic usage of patient portal software and lowers their security concerns  & \faMedrt \newline \faUsb \newline \faUsers \newline \faExclamationTriangle \newline \faMapPin \newline \faMapMarked \\
    \hline
    \cite{yin2021examining} & Evaluate users' perceptions and trust factors in patient portal software & Quantitative: logistic models &  Participants who value their portals for managing their healthcare are more likely to trust their portals. & \faMedrt \newline \faExclamationTriangle\\
    \hline
    \cite{o2019patient} & Evaluate Patients' perceptions of the risks and advantages of linking existing research data sources & Quantitative: descriptive statistics & $19.7\%$ of the participants are weary about researchers having access to their deidentified data.\newline $90\%$ of the participants are more assured when their unique identifiers were removed from the the dataset used for research and linkage & \faShareSquare	\newline \faUserFriends \newline \faExclamationTriangle\\
    \hline
    \cite{mancilla2009exploring} & Investigate admitting and registration protocols in hospital in order to establish best practices to curtail medical identity theft & Mixed Methods: descriptive statistics for quantitative data, thematic analysis for qualitative data & $78.5\%$ of the participants confirmed that patient identities is verified at admission or registration $91.9\%$ of which using driver's license. If the patient shows up without proof of identity, $59.5\%$ of the participants affirmed that they provide the service without confirming the identity of the patient & \faGavel \newline \faMedrt \newline \faExclamationCircle\\
    \hline
    \cite{coventry2020cyber} & Understand the insecure practices within healthcare  & Qualitative: thematic analysis & Three main impediments for security: security viewed as a barrier to patient care and productivity, Ignorance of consequences, dearth of policies and reinforcement of secure behaviour & \faGavel \newline \faExpeditedssl \newline  \newline \faExclamationCircle \newline \faExclamationTriangle\\
    \hline
    \cite{baker2011healthcare} & Understand security and privacy practices of physicians' offices' staff  & Qualitative: phenomenological approach & Several insecure behaviours were observed such as password sharing, data left in insecure areas and absence of password use & \faShareSquare \newline \faMedrt \newline \faExclamationCircle \newline \faMapPin\\
    \hline
    \cite{kozyreva2021psychological} & Evaluate the public's perceptions and acceptance of contact tracing technologies & Quantitative: descriptive statistics + logistic models + chi-squared tests & In March 2020, $68\%$ of participants declared that it was acceptable to grant the government access to citizens' medical records vs only $35\%$ participants in November of the same year  \newline Acceptance of privacy intrusive technologies diminished over time during the pandemic. & \faUsb \newline \faExclamationTriangle \newline \faMapMarked\\
    \hline
    \cite{deverka2019hopeful} & Investigate the public's perceptions about the important concerns in the design of medical information commons (MIC)  & Qualitative: thematic analysis & There needs to be a balance between the benefits of an MIC and the safeguards it implements to keep patients' data private &\faShareSquare \newline \faUsers \newline \faExclamationTriangle\\
    \hline
    \cite{adanijo2021investigating} & Analyse the outlook of the mental health service users on satisfactory data sharing practices & Qualitative: thematic analysis & Participants expressed concern over the security and the high risk of large datasets. \newline 
    Participants conveyed the  necessity to preserve the privacy and confidentiality of patients while taking into consideration the people who have access to privileged data. & \faShareSquare \newline \faExclamationTriangle \\
    \hline
    \cite{fylan2021co} & Investigate the participants' perceptions on healthcare data sharing process and establishing ways to gain their trust of the process & Participants expressed concerns over being identified and security limitations of data sharing systems \newline Participants declared that their primary care providers as well as hospital doctors and nurses should have access to their medical records & participants approve and advocate for sharing healthcare data for direct care, but not for social care. \newline Participants expressed concerns over privacy, security limitations and potentially having providers make biased decisions based on information found in their records &\faShareSquare \newline \faMedrt\\
    \hline
    \cite{schnall2015trust} & Examine the factors that contribute to patients' intention of using an HIV mobile healthcare application including security, privacy, trust, risk and usability& Qualitative: thematic analysis & Participants expressed concerns over privacy and trust of their sensitive healthcare data and the people who would have access to their healthcare data \newline Participants worried about the perceived risks including disclosure, tracking and data leaks & \faExclamationTriangle \newline \faMobile\\
    \hline
    \cite{ermakova2016improving} & Investigate how promises of confidentiality contribute to the participants' willingness to accept health clouds as an infrastructure for healthcare data sharing & Quantitative: descriptive statistics + Comparison of means & The promise of privacy increases the participants acceptance of health clouds in the case of sensitive and confidential healthcare data on the other hand, no statistical significance was found in the case of non-sensitive medical data & \faExpeditedssl \newline \faShareSquare \newline \faUsb\\
    \hline
    \cite{ramli2021hipaa} & Assess the understanding and healthcare data security awareness levels of participants& Quantitative: descriptive statistics & Participants' knowledge is lacking: (mean=2.6 where the average should be less than 2). Hospital management has the highest security awareness levels (mean=2.0667) while physicians have the lowest (mean=2.9202) &\faGavel \newline \faExclamationCircle\\
    \hline
    \cite{elger2009violations} & Assess the knowledge and perceptions of physicians on healthcare data violations of privacy and confidentiality & Quantitative: descriptive statistics + Comparison of means & Barely $11\%$ of the participants recognized all the confidentiality violations in the test cases they were presented with  & \faGavel \newline \faExpeditedssl \newline \faShareSquare \newline \faEyeSlash\\
    \hline
    \cite{tjora2005privacy} & Analyze the privacy posture of patients who use secure electronic communication systems (ECS) compared to their perception on usability of these systems & Qualitative: thematic analysis & Patients use the ECS for subjects they view as unsubstantial and avoid it for intimate or personal details &\faExpeditedssl \newline \faExclamationTriangle\\
    \hline
\caption{An Overview of the Security Focused User Studies Including Goal of each Study, Methods and Principal Findings. The symbols in the "Labels" column refer to the labels derived during the card sorting exercise: \faGavel = Regulatory Compliance, \faExpeditedssl = Secure Communication,\faShareSquare = Data Sharing, \faMedrt = EHR ,\faUserFriends = Individual Differences, \faExclamationCircle = Risk Awarness, \faUsb = Tech Adoption, \faUsers = Social Influence, \faExclamationTriangle = Risk Perception, \faMobile = Mobile Healthcare, \faEyeSlash = Privacy, \faMapPin = Contact tracing} \label{tab:long}
\end{longtable}
\end{small}
\end{center}

\end{document}